\pdfoutput=1
\documentclass[11pt]{article}

\usepackage{acl}

\usepackage{times}
\usepackage{latexsym}
\usepackage{tcolorbox}
\usepackage{fancyvrb}
\usepackage{xcolor}
\usepackage{listings}
\usepackage{paralist}

\usepackage[T1]{fontenc}
\usepackage[utf8]{inputenc}

\usepackage{microtype}

\usepackage{inconsolata}

\usepackage{graphicx}

\title{ETF: An \underline{E}ntity \underline{T}racing \underline{F}ramework for Hallucination Detection\\ in Code Summaries}

\author{Kishan Maharaj$^\dagger$, Vitobha Munigala$^\S$, Srikanth Tamilselvam$^\S$, Prince Kumar$^\S$,\\ \textbf{Sayandeep Sen$^\S$, Palanivel Kodeswaran$^\S$, Abhijit Mishra$^\ddagger$, Pushpak Bhattacharyya$^\dagger$}
\\
$\dagger$ Indian Institute of Technology Bombay, Mumbai, India \\
$\S$IBM Research India\\
$\ddagger$ University of Texas at Austin, Texas, United States\\
   \texttt{kishan.maharaj.iitb@gmail.com, \{vmunig10, srikanth.tamilselvam\}@in.ibm.com}\\ \texttt{prince.kumar12@ibm.com, \{sayandes, palani.kodeswaran\}@in.ibm.com}, \\ \texttt{abhijitmishra@utexas.edu,      
    pb@cse.iitb.ac.in}}

\begin{document}
\maketitle
\begin{abstract}
Recent advancements in large language models (LLMs) have significantly enhanced their ability to understand both natural language and code, driving their use in tasks like natural language-to-code (NL2Code) and code summarisation. However, LLMs are prone to hallucination—outputs that stray from intended meanings. Detecting hallucinations in code summarisation is especially difficult due to the complex interplay between programming and natural languages. We introduce a first-of-its-kind dataset, CodeSumEval, with $\sim$10K samples, curated specifically for hallucination detection in code summarisation. We further propose a novel Entity Tracing Framework (ETF) that a) utilises static program analysis to identify code entities from the program and b) uses LLMs to map and verify these entities and their intents within generated code summaries. Our experimental analysis demonstrates the framework’s effectiveness, leading to a 73\% F1 score.  The proposed approach provides a method for detecting hallucinations by tracing entities from the summary to the code, allowing us to evaluate summary accuracy and localise the error within the summary. 

\end{abstract}

\section{Introduction}

Hallucination in natural language processing is defined as a condition in which a language model produces a text that is either incoherent or does not faithfully represent the provided source input \cite{ji2023survey}. Similarly, in the context of code summarisation, \textit{hallucination can be defined as a condition in which the generated summary does not accurately capture the intent and implementation details of the given input code.} 

\begin{tcolorbox}[,
    colback=gray!10,     % Background color
    colframe=gray!30,    % Border color
    coltitle=black,
    fonttitle=\bfseries, % Font style for the title
    top=0mm,
    left=0mm,            % Left margin
    right=0mm,           % Right margin
    bottom=-2mm-2mm,
    float=ht!
]
\small
\begin{lstlisting}
public RowBuilder int16(String name)
    {
    ColumnInt16 column = new ColumnInt16(_columns.size(), name, _offset);
    _offset += column.length();
    _columns.add(column);   
    return this;
 }
\end{lstlisting}

\textbf{Summary}:
.......
This method is used to add a new column of \textcolor{red}{data type int16 (16-bit integer) } to the existing data structure. It creates a new ColumnInt16 object with the given name and size (16 bits)...........
\tcbsubtitle[halign=center]{Example 1 (LLama3-70B): Confused Data Type}

\end{tcolorbox} 
% \captionof{figure}{"LLama3-70B Summary: Confused data type example 1 example 2"}
\label{box: int16_example_top}

% Detecting hallucinations in code summaries is particularly challenging due to several factors: (a) the naturalness of the generated content, (b) the complex interplay of code snippets within the summary, and (c) the presence of a high degree of polysemy. 
\begin{figure*}
  \centering
  \includegraphics[width=\textwidth]{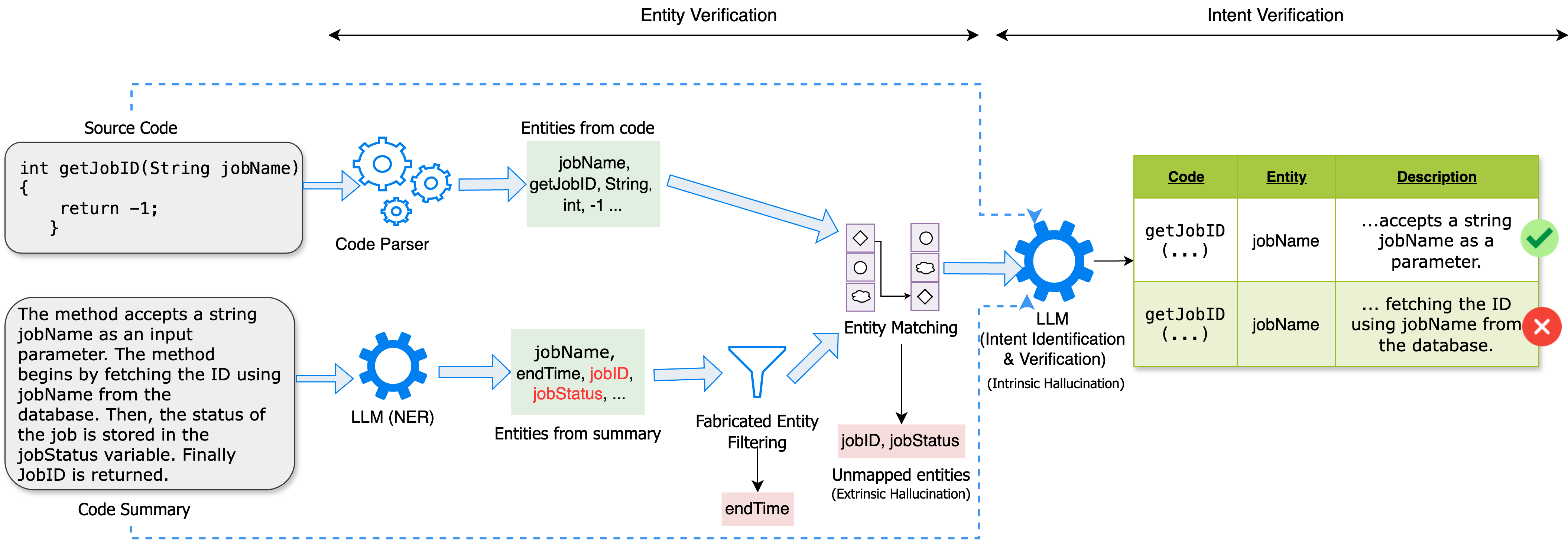}
  
  \caption{Proposed Methodology: This diagram illustrates our end-to-end Entity Tracing Framework (ETF), which takes source code and a corresponding summary as input and returns whether the summary is hallucinated or not. First, we use code parsers to extract entities from the source code and employ large language models (LLMs) to identify entities from the summary. Next, we apply string-based heuristics to match entities from the summary to the code. Following this, an LLM verifies the accuracy of each entity’s description by cross-referencing the source code with relevant sentences in the summary. This process enables the localisation of hallucinated content in the summary, ultimately enhancing its explainability.}
\end{figure*}
\label{fig:sys-arch}

% Hallucination can originate from a combination of factors, including limited understanding of programming functionalities, code complexity, and misinterpretation of code entities.
Hallucination in code summaries often stems from misinterpreting entities, complex logic, or limited model understanding. This can impact the ability of models to interpret the intended functionality of the code, resulting in an inaccurate portrayal of its purpose. With the recent advances in language models \cite{achiam2023gpt, touvron2023llama, mishra2024granite, team2023gemini}, the hallucination detection in code summarisation can become very challenging due to the tendency of models to produce convincing but inaccurate summaries. For instance, consider the Example \ref{box: int16_example_top}, where the intention of the \textit{int16} java method is to create a new 16-bit integer column (ColumnInt16) with a specified name, update the position for the next column, add it to the list of columns, and then return the RowBuilder object. However, the generated explanation introduces a non-existent \textit{int16} datatype and proceeds to discuss the rest of the logic as if it were valid. It can be noted that \textit{int16} is a valid datatype in other programming languages such as C, C++, C\#, and Go, causing LLMs to confuse it with learning from those languages. In general, these statements could also mislead a novice Java developer into believing that an \textit{int16} datatype exists in Java. Furthermore, several large language models (LLMs) like Llama and Granite failed to detect this hallucination. 
Similarly, in the example shown in Figure \ref{fig:sys-arch}, the Java method getJobID() takes ``jobName" as an argument and simply returns -1.
The summary generated by the model provides a detailed explanation, including how the method getJobID() connects to the database and attempts to retrieve the jobID using the given jobName. Additionally, the summary mentions it stores the ``jobStatus" in a variable. Clearly, the generated summary has no supporting entities in the code for database access, and the model is relying on the method name to hallucinate a plausible summary of the method.

We focus on detecting hallucinations in code summaries, cases where generated text misrepresents the code. Our study is motivated by the growing real-world use of LLM-based code documentation tools like Amazon Q Developer\footnote{\href{https://aws.amazon.com/q/developer/}{Amazon Q Developer}} and IBM Watsonx Code Assistant\footnote{\href{https://www.ibm.com/products/watsonx-code-assistant}{IBM Watsonx Code Assistant}}. Therefore, detecting hallucinations in code-related outputs is critical to ensuring the reliability and correctness of these systems in general software engineering workflows.

We study different factors that can lead to hallucination and list down a taxonomy to map the common causes easily. Noting a lack of datasets to reliably research this topic, we create a first-of-its-kind dataset, \textbf{CodeSumEval}, for studying hallucination in code summarisation with \textbf{411 summaries} generated by seven different large language models, broken into \textbf{9933 entity-level samples}. This dataset consists of code and a corresponding summary describing the code. The code summaries are generated with the original CodeXGLUE \cite{lu2021codexglue} dataset, which consists of code snippets scraped from open-source GitHub repositories.  All code snippets used are unaltered and authentic, thereby reflecting realistic, real-world usage scenarios. The annotation consists of a) NER, b) Entity Description Verification, and (c) Overall Summary Quality (not focusing on completeness or conciseness).  Importantly, all the hallucination cases discussed in the paper are naturally occurring in code summaries from the models and are not synthetically constructed or perturbed, thereby incorporating the real-world challenges of convincing summaries.

Further, we introduce a framework that evaluates the correctness of the generated summary. For this, we verify if the entities discussed in the summary are present in the code and correctly described in the summary. The framework leverages code parsers like javalang\footnote{\href{https://github.com/c2nes/javalang}{javalang Repository}} to list the different entities in the code snippet and prompt-based approaches to detect entities in the summary. We note that detecting entities in the generated summary is more difficult due to the high degree of polysemy \cite{tabassum2020code}. For example, entities like "list", "while", "if", etc, can be code entities or natural language entities. This necessitates our reliance on large language models with high reasoning capabilities for detecting entities on the summary side. We then map the detected entities from the summary to code by using string-matching heuristics. The sentences with unmapped entities can be considered as unground (source of extrinsic hallucination). For each mapped entity, we have a tuple \textit{<code, entity, intent-related sentence>}, where the intent-related sentence can be considered as the sentence in summary mentioning the entity. Finally, we verify each tuple from the summary for intrinsic hallucination to assess the correctness of the code summary. Our experiments demonstrate the importance of localising entities in the summary for effective hallucination detection.
Our contributions are:
\begin{itemize}
    \item A taxonomy covering diverse reasons that might lead to hallucination in the code summarisation (Figure \ref{fig:taxonomy-code-hallucination}).
    
    \item A novel dataset CodeSumEval\footnote{\href{https://github.com/kishanmaharaj/hallucination-in-code-summarisation}{GitHub Link}} for studying hallucination detection in code summarisation, featuring 411 summaries from 7 LLMs and $\sim$ 10K entity-level samples (Table \ref{tab:Data_Statistics}) with an explanation of causes for hallucinations as per taxonomy.
    
    \item A first-of-its-kind approach for hallucination detection in code summarisation inspired by the insights from human behaviour during code reviews, leading to a performance of 73\% F1 score (Table \ref{tab<code_hallucination>:hallucination_detection}).
    
\end{itemize}

\section{Related Work}
% \textbf{Kishan; Vitobha; Prince}

Recent advances in the NLP community have witnessed significant improvements in hallucination detection pipelines. In this section, we discuss some of the works that are relevant to ours.  

\textbf{Hallucination in Natural Language}: \citet{rawte2024tutorial, sahoo2024addressing} review recent advances in hallucination detection in natural language, emphasising its practical significance.  Recently, prompt-based methods (\citet{arora2022ask}, \citet{manakul2023selfcheckgpt}, \citet{agrawal2023language}, \citet{dhuliawala2023chain}) are being used to detect hallucinations in the text produced by LLMs. 
% The self-consistency framework inspires the central idea of such approaches and suggests that if an LLM knows a given concept, sampled responses are likely to be similar and contain consistent facts, while hallucinated responses, when stochastically sampled, are likely to diverge and contradict one another.
\citet{xiao2022entity} and \citet{zhang2022improving} attempt to address entity-level verification in natural language inputs. Both of these works involve improving the correctness of natural language summaries and do not discuss anything in the context of code. We note that most of the hallucination detection frameworks \cite{manakul2023selfcheckgpt, arora2022ask, dhuliawala2023chain, valentin2024cost, rebedea2023nemo}  in natural language do not enforce reference text for grounding. In our setup of code summarisation, the generated summary has to be evaluated with respect to a reference text (the code snippet). Therefore, necessitating an approach which could compare the code summary to the code snippet. \citealp{maynez2020faithfulness, ji2023survey} discuss further fine-graining of hallucination in natural language as intrinsic and extrinsic hallucination. More specifically, \textbf{Intrinsic hallucination} occurs when the given text contradicts the reference, while \textbf{Extrinsic hallucination} happens when the text cannot be verified against the reference. We use a similar convention in our paper.

% More specifically, \citet{xiao2022entity} introduces a scan-copy mechanism which involves correcting the factuality of the decoder output by copying the entities directly from the documents. While \citet{zhang2022improving} aims to improve the correctness of the summary by controlling the coverage of entities in the summary at inference time. 

\textbf{Hallucination in Code Generation:} The code generation space has captured significant attention due to its practical significance in software development. \cite{jiang2024survey} discusses recent developments in code generation and suggests the importance of addressing hallucination for improving the reliability of LLMs.  \citet{liu2024exploring} studies hallucination in code generation and proposes a categorisation that encompasses five categories of hallucinations based on the conflicting objectives and varying degrees of deviation observed in code generation. \citealp{tian2024codehalu, agarwal2024codemirage, spracklen2024we} advanced the field with datasets and frameworks addressing hallucination in code generation. These studies highlight that while LLM-generated code may be syntactically correct and semantically plausible, it often fails to execute as intended or meet requirements.

Despite progress in hallucination detection, code summarisation remains underexplored. \citealp{kang2024identifying} and \citealp{zhang2024detecting} focused on inconsistencies in comment generation, addressing specific aspects like design constraints and parameter types, but their methods face challenges due to reliance on execution environments. In contrast, our approach validates the full functionality of generated outputs, independent of external dependencies, offering a more reliable solution by grounding entities in the input code and verifying their intent.

% further advanced the field by proposing valuable datasets and frameworks to tackle hallucination in code generation. These studies indicate the tendencies of LLMs to produce code that is syntactically correct and even semantically reasonable but fails to execute as intended or meet the specified requirements. 

% Although there have been significant advances in hallucination detection in code generation, the field of code summarization is still nascent. Recently \citealp{kang2024identifying, zhang2024detecting} have investigated inconsistencies in comment generation, but their focus has been limited to specific aspects such as verifying design constraints like parameter types and ranges. Additionally, some of their approaches, such as generating test cases for comments, face practical challenges due to heavy reliance on execution environments. In contrast, our approach aims to validate the entire functionality described in the output, independent of external environments or dependencies. By grounding entities based on the input code and verifying their intent, our framework offers a more reliable and less dependent solution.

% To the best of our knowledge, no prior study has been done to address hallucination detection in code summarization.    

% Other points to be mentioned:
% \begin{itemize}
%     \item Stackoverflow NER
%     \item code generation hallucination
%     \item promt based hallucination detection
%     \item Errors in code summaries
% \end{itemize}
\section{CodeSumEval Dataset}
% \textbf{Kishan; Prince; Vitobha}

% \begin{itemize}
%     \item CodeXglue 
%     \item filtering the codexglue part
%     \item annotation process
%     \item final stats
% \end{itemize}

To create the CodeSumEval dataset, we consider code snippets from the Java programming language and CodeXGLUE \cite{lu2021codexglue} -- Code-To-Text dataset. We focused on the Java programming language due to its widespread relevance in the industry. It offers a rich set of entities (such as classes, methods, and variables) due to its structured design and strict typing system. The dataset was annotated by 8 annotators who are experts in Java and hold at least a Master’s degree in Computer Science, with some having a PhD in the field. On average, the annotators had 4+ years of experience in Java programming. We report the statistics in Table \ref{tab:Data_Statistics} and describe the data curation process below:

% To ensure the variations in the lengths, we created three different bins based on code token counts: Small, consisting of 25 to 50 code tokens; Medium, consisting of 50 to 100 code tokens; and Large, consisting of 100 to 200 code tokens. We ensure a balanced distribution by including 200 instances from each length category.

% Mention a Diagram and define code tokens 
% Mapping the models to code taxonomy 
% Figure 2 to the Page 1
\textbf{Summary Generation}: 
% The code descriptions in the CodeXGLUE dataset are usually a few lines of comments and do not capture many important details. Therefore, 
We generate summaries from CodeXGLUE by prompting seven different LLMs (Appendix \ref{box: summary generation prompt}) with 600 code snippets. By producing multiple summary variants, we can assess hallucination generation by different LLMs and evaluate hallucination detection techniques under varied conditions. We present quantitative results (Table \ref{tab<code_hallucination>:Quantitative_Analysis}) and qualitative analysis in Section \ref{sec:analysis}. During initial annotation, we found that annotators spent considerable time verifying summaries, often requiring online documentation searches, leading to an average annotation time of 30 minutes or more per summary. To ensure feasibility, we randomly prune samples and use $\sim$10\% of the data for the final hallucination annotation task (Table \ref{tab:Data_Statistics}).

%The given prompt comprises three different aspects of the code summary involving  1) Input/Output of the method describing the input parameter of the method and the return value of the method; 2) Business Purpose describing a high-level functionality of the given code and 3) Detailed functional summary of the method describing the flow of the code in a detailed manner.

\begin{table}[h!]
\centering
\begin{tabular}{l|cc}

\textbf{Category}    & \textbf{Count}   & \textbf{Percentage (\%)} \\
\hline
\multicolumn{3}{c}{\textbf{Summary Level Classification}} \\
\hline
Hallucinated          & 130              & 31.63\%               \\
Not Hallucinated      & 281             & 68.36\%               \\
\hline

\textbf{Total Summaries} & \textbf{411}  & \textbf{100\%}        \\
\hline
\multicolumn{3}{c}{\textbf{Entity Level Classification}} \\
\hline
CORRECT           & 9024           & 90.84\%               \\
INCORRECT         & 303            & 3.05\%                \\
IRRELEVANT         & 606            & 6.11\%                \\
% IRRELEVANT        & 137            & 1.70\%                \\
\hline
\textbf{Total Entities} & \textbf{9933}  & \textbf{100\%}        \\
\hline
\end{tabular}
\caption{Overall Data Statistics}
\label{tab:Data_Statistics}
\end{table}
\textbf{Named Entity Recognition} Since our framework involves tracing entities from summary to code, we perform NER of the summaries based on the tagset suggested in \citet{tabassum2020code} ( prompt in Appendix \ref{box: named entity recognition prompt}). 
    % We use the same model that generated the summary for detecting the entities in the summary. 
    % Human annotators were asked to validate and correct the entities and the types detected by these models to create the ground truth.
    %Specifically, the code entities include mentions of CLASS, VARIABLE, IN LINE CODE, FUNCTION, LIBRARY, VALUE, DATA TYPE, and HTML XML TAG. Whereas, the natural language entities include mentions of APPLICATION, UI ELEMENT, LANGUAGE, DATA STRUCTURE, ALGORITHM, FILE TYPE, FILE NAME, VERSION, DEVICE, OS, WEBSITE, and USER NAME. 

    % \item \textbf{Code Parsing:} We rely on static program analysis to parse the code and obtain the type of code entity. Specifically, we consider JavaLang parser\footnote{https://github.com/c2nes/javalang} for recognizing the code entities from the code snippets. We evaluate the named entity types based on this program analysis method by considering the output obtained from Javalang as ground truth.

\textbf{Hallucination Labelling:} For each detected code entity in a summary, all sentences describing that entity are considered relevant. To account for the scenario where the relevant sentence can be noisy, we introduced a third label, "IRRELEVANT". These instances can be used to evaluate the performance of the intent-detection module and removed during the preprocessing. Thus, we obtain tuples of (code, entity, relevant sentences) for each entity. A total of 9933 such tuples, sampled from 441 summaries, were selected for human annotation (hired based on volunteering) to detect hallucinations. Out of these, 4354 tuples (from 222 summaries) were independently reviewed by two different sets of annotators, leading to a Cohen Kappa score of 0.72, implying high agreement. The conflicts were resolved by two independent meta-annotators. The annotators were asked to evaluate the overall summary by assigning a label of `GOOD', `FAIR', and `POOR'. We observe that, on average, 1.33 entities were marked as hallucinated for the summaries rated as `FAIR' or `POOR' \label{<content>:threshold}. Therefore, we consider a summary as hallucinated if at least one of the entities is hallucinated. After preprocessing, we consider the instance with labels "CORRECT" and "INCORRECT" in human data and treat the "IRRELEVANT" label predicted by the model as "INCORRECT". We provide the complete annotation guideline in Appendix \ref{appendix:annotation_guidelines}.

    % We also leveraged annotation from GPT4-Omni for increasing the data size and observed moderate agreement between the annotation of GPT and humans of 0.534. Based on our qualitative analysis, we conjecture that this disagreement may due to the different annotation style of GPT compared to humans. We elaborate on these points in the results section.

    % \textcolor{red}{@Kishan - We can add a dataset statistics table here. Number of <code, summary samples>, total hallucinated samples, avg entities, etc.}

    %\item \textbf{Hallucination Labelling: \textcolor{red}{To be rewritten considering human annotation}} For the annotation of hallucination, we leverage GPT-4o due to its superior performance on code-related benchmarks like HaluEval. It is worth noting that the performance metrics are very close to human performance. 
    %For studying hallucination in the context of code summarization, it is important to note the high sensitivity of the correctness in terms of code execution i.e., the code may not execute correctly even if a single line is incorrectly written in the code. Therefore, to construct the code back from the summary, every line of the summary must be correct. 
    %We characterize an instance of summary as hallucinated if the summary consists of one or more lines of incorrect description of the given input code. 

\section{Categorisation of Factors for Hallucination in Code Summarisation}

\label{sec:factors-contributing-to-hallucination-in-code-summaries}

\begin{figure*}
  \centering
  \includegraphics[width=\textwidth]{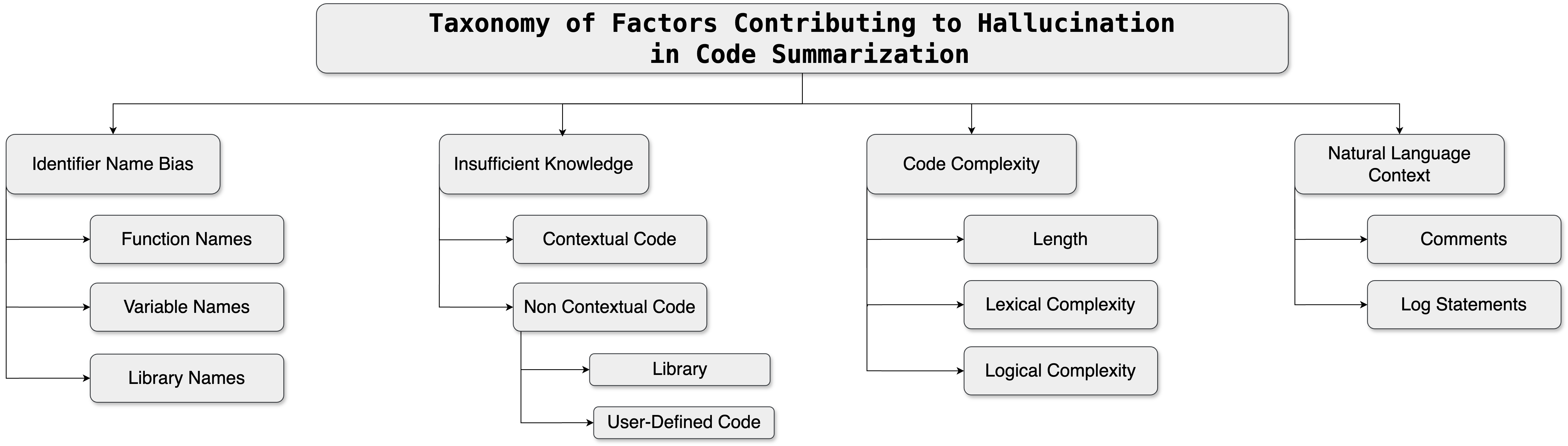}
  \caption{Taxonomy of hallucination in code summarisation based on the causes of hallucination. We start with four broad categories and then present a fine-grained classification of each category.}
  \label{fig:taxonomy-code-hallucination}
\end{figure*}
In this section, we describe the various factors that could lead to hallucination in code summaries (Figure \ref{fig:taxonomy-code-hallucination}) based on what we learned from the annotation process. This classification, based on the underlying factors of hallucination, offers insights into the generative behaviours of language and code models. We describe these categories of hallucination factors below and discuss their statistical analysis in Figure \ref{fig:hallucination_distribution}.

\label{sec:HC1-Name_Bias}
\textbf{HC1: Based on Identifier Name Bias}: Name Bias refers to the tendency of language models to rely on identifier names when interpreting code. We classify this bias into three subcategories based on its source: 1)Variables, 2)Functions, and 3)Libraries. The model can misinterpret code due to the linguistic characteristics of these entity names. As the semantics of the code is defined by the underlying logic rather than their lexical meaning of entities, this may lead to hallucination. In the example shown in Figure \ref{fig:sys-arch}, the model (Granite-20B)
incorrectly assumes that \texttt{getJobID} is about
retrieving a job ID, based purely on their names, even though the actual code logic suggests otherwise.

\textbf{HC2: Insufficient knowledge}: This involves scenarios where the model generates incorrect summaries due to the lack of knowledge. This may include an incorrect explanation of the imported libraries that the model did not see in its training data, incorrect information about the keyword, etc. We further divide this category into two parts:

1) \textbf{Contextual code:} This occurs when the model fails to correctly explain the code, often because it has not encountered the functionality of code during training or is working with a low-resource language like COBOL, where fundamental rules may be misrepresented in the summary.
% b) \textbf{Incorrect Reasoning} pertaining to the scenario when the model has insufficient reasoning prowess to understand the given code correctly.

2) \textbf{Non-contextual code:} This involves the scenario when the input does not contain the complete code and mentions an unseen library or an unknown construct whose functionalities are not understood by the model. For example, in the code sample shown in the \hyperref[box: HC2-example]{HC2 Example}, the model incorrectly describes the purpose of SQLException.

\begin{tcolorbox}[,
    colback=gray!10,     % Background color
    colframe=gray!30,    % Border color
    coltitle=black,
    fonttitle=\bfseries, % Font style for the title
    top=0mm,
    left=0.1cm,            % Left margin
    right=0.1cm,           % Right margin
    bottom=-1mm-0.5mm,
]
\small
\begin{lstlisting}
public String getString (int columnPosition) throws 
SQLException {
    return (String) resultSet.getObject(columnPosition);
    }
\end{lstlisting}

\textbf{Summary}:
...The method first checks if the ResultSet object is null. \textcolor{red}{If it is, a SQLException is thrown}….
\tcbsubtitle[halign=center]{HC2: Granite-20B insufficient knowledge}
\end{tcolorbox}
\label{box: HC2-example}

\textbf{HC3: Code Complexity}: 
This pertains to the model's tendency to produce incorrect code summaries due to high code complexity. This may stem from the model's insufficient reasoning capabilities to understand the code or the misinterpretation of user instructions. Key factors contributing to complexity include: 1) \textbf{Length}: Longer code is more complex and involves more interdependencies with more potential points of failure. 2) \textbf{Lexical Complexity}: Complex vocabulary, including diverse operands and operations, increases the number of elements to track and understand. 3) \textbf{Logical Complexity}: Code with high cyclomatic complexity, multiple paths, or distant method invocations increases the challenge of comprehension.
% This category highlights the model's tendency to produce incorrect code summaries due to high code complexity. From another perspective, the model may not have enough reasoning capabilities to understand the code correctly and may deviate from the instruction given by the user. These complexities can arise due to various reasons, such as 1) \textbf{Length}: Longer code tends to be more complex because it often requires more understanding, has more potential points of failure, and may involve more interdependencies. 2) \textbf{Lexical Complexity}: This refers to scenarios where complex vocabulary in code, including diverse operands (variables, constants) and operations (functions, operators), increases complexity due to the higher number of elements to track and understand. 3)  \textbf{Logical complexity}: This refers to scenarios where the code's logic is complex, often due to high cyclomatic complexity, indicating many independent paths. Complexity increases with numerous paths or method invocations, especially when these are not directly visible or are called from distant parts of the codebase.
% This includes the scenarios where the model produces summaries unrelated to the code or unrelated junk codes, etc.

% \captionof{figure}{}

\begin{tcolorbox}[,
    colback=gray!10,     % Background color
    colframe=gray!30,    % Border color
    coltitle=black,
    fonttitle=\bfseries, % Font style for the title
    top=0mm,
    left=0cm,            % Left margin
    right=0cm,           % Right margin
    bottom=-1mm-0.5mm,
    float=ht!
]
\small
\begin{lstlisting}
public <V> V execute(RedisCallback<V> cb) {
  Jedis jedis=jedisPool.getResource();
  boolean success = true;
  try {
    return cb.execute(jedis);
    } catch (JedisException e) {
        success = false;
        if (jedis != null) {
            jedisPool.returnBrokenResource(jedis);
        } throw e;
    } finally {
        if (success) {
            jedisPool.returnResource(jedis);
        }
    }
}
\end{lstlisting}

\textbf{Summary}:
...it then returns the `jedis" object to the pool using the `returnBrokenResource" method \textcolor{red}{if the `success" variable is `false"}...
\tcbsubtitle[halign=center]{HC3: Codellama-7B confusion with conditions}
\end{tcolorbox} 
\label{box: HC3-complexity example}

In the code snippet shown in \hyperref[box: HC3-complexity example]{HC3 Example}, the model Codellama-7b \cite{roziere2023code} produces an incorrect interpretation of the condition. This may be due to increased complexity due to nesting, leading to more complicated logic that is challenging for the model to understand.

\begin{tcolorbox}[,
    colback=gray!10,     % Background color
    colframe=gray!30,    % Border color
    coltitle=black,
    fonttitle=\bfseries, % Font style for the title
    top=0mm,
    left=0.1cm,            % Left margin
    right=0.1cm,           % Right margin
    bottom=-1.5mm-0.5mm,
    float=hb!
]
\small
\begin{lstlisting}
public static HashSet<String> createSetFromProperty(String property) {...
    if (property != null && !property.equals("null")) { 
    // "([\\w]*)=([\\w]*);"
    Pattern params =     Pattern.compile("([\\w]+)[;]*");
    ...}
\end{lstlisting}

\textbf{Summary}:
...
The input string is expected to contain a list of properties in a specific format, where \textcolor{red}{each property consists of a name-value pair (e.g., "name=value;")}...
\tcbsubtitle[halign=center]{HC4: LLama3-70B mislead by the comment}
\end{tcolorbox} 
\label{box: HC4-comment example}

\textbf{HC4: Natural language Context}: This refers to cases where natural language in code snippets, such as outdated \textbf{comments} or \textbf{log statements}, causes hallucinations in code summaries. In the code snippet shown in \hyperref[box: HC4-comment example]{HC4 Example}, the LLama3-70B model incorrectly infers that the property variable contains a list of key-value pairs inferred from a commented line. However, the `property" variable contains an alphanumeric string followed by one or more semicolons.

% \captionof{figure}{}

% For example, in the above code, the mentioned comments discuss the earlier functionality of the code and act as a misinformation source for the code summary.  

\section{Methodology}
% \textbf{Kishan; Vitobha}

%% Elaborate on how a detailed summary of a code consists of entity and actions 

A code summary typically has a global and local view similar to texts \cite{maharaj2023eyes}. While the global view includes purpose, functionality, control flow, data flow, etc., the local view includes the details of key entities (variables, functions, etc.) from the source code and their purpose (hereby referred to as the intent of the entity). Our approach is based on the intuition that software developers, while verifying the documentation for a given code repository, first understand the local aspects of the code and then build a bottom-up concept for understanding the global aspects of the code. This involves reading the code line by line and tracing the specific code entities from the documentation to the original code. 

This behaviour aligns with working memory theory in cognitive science \cite{baddeley1994developments}; working memory is a brain system that temporarily stores and manipulates the information necessary for complex cognitive tasks like learning and reasoning. The capacity of working memory is bounded by $7\pm2$ objects at any point in time, which further reduces to 3-4 objects \cite{cowan2014working} if the objects have relational dependencies with each other. Since code summaries often involve interdependent objects, developers must focus on local aspects to build a global understanding, suggesting a bottom-up heuristic for code summary comprehension. We leverage these behavioural insights to design an LLM-powered framework for detecting hallucinations in code summaries, which involves tracing the entities from the summary to the code. This aspect of mapping the entities from the summary to the code aims to simulate the bottom-up behavioural model of verifying the description of coding entities at a time. With these insights, we aim to measure the correctness of a code summary as a two-step process: (1) Entity Verification and (2) Entity-Intent Verification. The detailed flow of this framework can be found in Figure \ref{fig:sys-arch}.

\subsection{Entity Verification}
In entity verification, we check if the entities in the summary are present in the source code to detect extrinsic hallucination. This involves extracting entities from both the code and summary, then mapping entities from the summary to the code. We elaborate on this process below:

\textbf{Entity Extraction from code}: We leverage program analysis to extract entities from code ({\it Javalang} Python package\footnote{\href{https://github.com/c2nes/javalang}{javalang Repository}}). The code is tokenised (lexer) and parsed into an abstract syntax tree (AST). This tree structure represents the hierarchical organisation of code elements, making it easier to analyse. This yields a fine-grained classification of all the tokens present in the code, such as variable names, class names, function names, etc.   
% along with their byte position relative to beginning of the code fragment. 

\textbf{Entity Extraction from summary}: 
% Named Entity Recognition (NER) is a common NLP task that extracts and categorizes entities from text (e.g., Organization, Person, Language). However, these generic tags are not directly applicable to code summaries. 
% Code summaries often focus on properties like Variables, Classes, Functions, and Tables, which cannot be categorized using traditional NLP tagsets. To address this,
\citet{tabassum2020code} proposes the task of entity detection in code summaries and introduces a relevant NER tagset. We adopt this tagset for extracting entities from code summaries (Prompt: Appendix \ref{Appendix<Prompts>} Figure \ref{box: named entity recognition prompt}). 
% and benchmark various LLMs in addition to the approach proposed in \cite{tabassum2020code}
% To perform NER using LLMs, we provide the code summary and NER tagset in the prompt (Appendix \ref{box: named entity recognition prompt}) using a one-shot in-context example to extract all the entities discussed in the summary accompanied by their types. 
Leveraging LLMs to recognise entities introduces the risk of hallucinations, where the model may fabricate entities not present in the code summary. To address this, we implement a filtration step to remove such fabricated entities. We evaluate Gemini and GPT-4-Omni for Code NER using human-collected data, with results in Appendix \ref{Appendix<ner_evaluation>}. We also assess all the open-source models considered in this study. Our findings show a strong correlation between GPT-4-Omni predictions and human data, confirming its effectiveness for entity detection in our framework.

\textbf{Entity Matching}: Once the entities from the code and summary are extracted, we compare them to identify the subset of entities present in the summary but not in the code. These entities are termed ungrounded, and all the sentences in the summary containing these entities can be labelled as extrinsic hallucination. The subset of entities in both the summary and the code goes through an additional verification round for intrinsic hallucination. This is to validate if the intent of the entity in the summary is correctly described as per the code.

\subsection{Entity-Intent Verification}
The presence of an entity in both the summary and code indicates that the entity is valid, but does not warrant the correctness of the context in which it is discussed. For example, in Figure \ref{fig:sys-arch} \textit{jobId} is a correct entity, but the context of \textit{retrieving jobID from the database} is incorrect. To address this problem, we propose verifying whether the intent of each mapped entity is accurately described in the summary. We extract all sentences containing the entity of interest from the summary to form its intent context. To identify these relevant sentences that describe an entity's intent, we explored two approaches: (1) prompt-based and (2) string-matching heuristics. Our qualitative assessment, detailed in Appendix \ref{app:intent_detection_details}, demonstrates that rule-based heuristics were both more effective and efficient than prompt-based methods, which were prone to hallucinations. Therefore, we relied on string-matching heuristics for our framework. After identifying the entity and intent, we use LLMs with zero-shot prompting to verify their correctness with the code (Prompt: Appendix \ref{box: Intent Verification prompt}). We also experimented with few-shot prompts by including examples of various hallucination types in code summaries along with the representative code. However, performance degraded due to the increased prompt length, consistent with findings in recent works like \citet{mirzadeh2024gsm}.

To identify the quality of the whole summary with respect to the code, we aggregate the individual entity-intent hallucination and set the threshold for labelling as 1, as discussed in (Section \ref{<content>:threshold}) following human annotation where a summary was rated as 'FAIR' or 'POOR' when an average of 1.33 entities were wrongly described. 
% Specifically, the summary is marked as hallucinated if at least one intent is identified as INCORRECT.

\begin{figure}
  \centering
  \includegraphics[width=0.5\textwidth, trim=0cm 0cm 0cm 2cm]{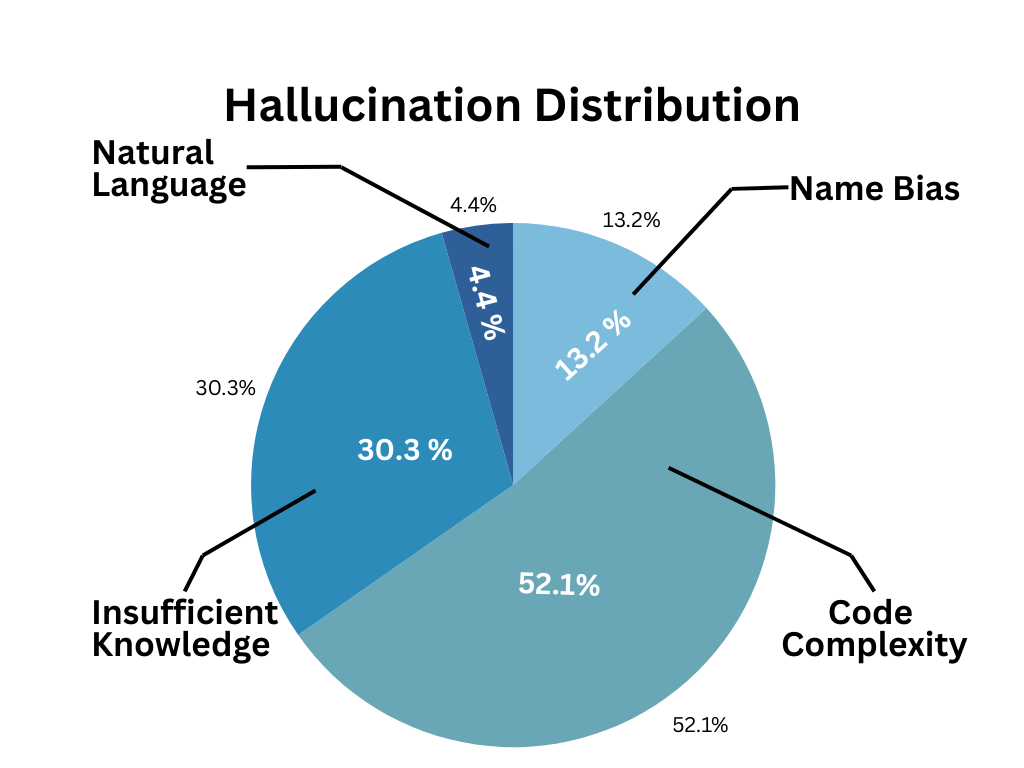}
  
  \caption{Distribution of different hallucination categories proposed in the taxonomy. We observe that the models tend to hallucinate most frequently due to the high complexity of the code, while significant instances of insufficient knowledge were also identified.}
\end{figure}
\label{fig:hallucination_distribution}

\section{Experiments and Results}
% \textbf{Kishan; Prince; Vitobha}
\begin{table}[ht!]
\centering
\begin{tabular}{|l|c|c|c|}
\hline

\textbf{Model}         & \textbf{P} & \textbf{R} & \textbf{F1} \\ \hline
\multicolumn{4}{|c|}{\textbf{Instance Level}}             \\ \hline
Gemini-2.0-Direct       & 0.51               & 0.50            & 0.42           \\ 
Gemini-2.0-ETF*       & 0.64               & 0.65            & 0.64               \\ \hline

% Gemini-1.5-Direct       & 0.41               & 0.49            & 0.25           \\ 
% Gemini-1.5-ETF*       & 0.68               & 0.64            & 0.51               \\ \hline

GPT4-Omni-Direct              & 0.48               & 0.50            & 0.28   \\

\textbf{GPT4-Omni-ETF* }             & \textbf{0.72}               & \textbf{0.74}            & \textbf{0.73}            \\ \hline

Mixtral-8x22B-Direct              & 0.48               & 0.48            & 0.45   \\

Mixtral-8x22B-ETF*              & 0.62               & 0.61      & 0.61            \\ \hline

Llama-3.1-70B-Direct              & 0.57               & 0.54            & 0.38      \\

Llama-3.1-70B-ETF*              & 0.62               & 0.62            & 0.54            \\ \hline

Llama3-8B-Direct             & 0.59               & 0.51            & 0.26            \\

Llama3-8B-ETF*              & 0.51               & 0.55            & 0.50            \\ \hline

Mistral-7Bv3-Direct              & 0.16               & 0.50            & 0.24   \\

Mistral-7Bv3-ETF*              & 0.51               & 0.50            & 0.41            \\ \hline

\multicolumn{4}{|c|}{\textbf{Entity Level}}         \\ \hline
Gemini-2.0       & 0.58               & 0.62            & 0.60                   \\ \hline

% Gemini-1.5       & 0.55               & 0.70            & 0.55                   \\ \hline
\textbf{GPT4-Omni}              & \textbf{0.59 }              & \textbf{0.69}            & \textbf{0.61}        \\ \hline

Mixtral-8x22B              & 0.48               & 0.38      & 0.39            \\ \hline

Llama-3.1-70B & 0.55               & 0.62            & 0.56                   \\ \hline
Llama3-8B   & 0.60               & 0.59            & 0.48   \\ \hline            
Mistral-7B             & 0.52               & 0.59            & 0.49            \\ \hline

% \multicolumn{5}{|c|}{\textbf{Entity Level 3 way}}         \\ \hline
% Gemini-1.5-Flash       & 0.44               & 0.48            & 0.39        & 0.80              \\ \hline
% GPT4-Omni              & 0.46               & 0.47            & 0.46        & 0.87              \\ \hline
\end{tabular}
\caption{ We report macro Precision (P), Recall (R) and F1. We consider two evaluation aspects: 1) Instance Level, which aims to label the entire summary, and 2) Entity Level, which labels individual entities in summaries. Here, * represents the proposed approach. 
}
\label{tab<code_hallucination>:hallucination_detection}
\end{table}

For summary generation, we consider instruction-tuned versions of the IBM-Granite family (20B and 34B) \cite{mishra2024granite}, Llama3 family (8B and 70B) \cite{touvron2023llama}, CodeLlama family (7B and 34B) \cite{roziere2023code} and Mistral-7B \cite{jiang2023mistral}.
% For intent verification, the use of LLMs to verify their own generation may be challenging due to inherent confirmation bias \cite{xie2023adaptive, feng2024don}, which enables the model to agree with their own generations. To avoid confirmation bias, we chose GPT4-Omni \cite{achiam2023gpt} and Gemini-1.5-Flash \cite{team2023gemini} for our entity-intent verification task. 
% For intent verification, we started our initial experiments with powerful open-source models like LLama3-70B \cite{touvron2023llama} but observed worse than random performance (F1 score- 0.37) due to the very high complexity of the task. Similar qualitative studies with other models in our preliminary experiments prompted us to explore advanced models with stronger capabilities in code reasoning, such as GPT4-Omni \cite{achiam2023gpt} and Gemini-1.5-Flash \cite{team2023gemini}.

For intent verification, we consider Llama3.1-70B, Llama3-8B \cite{touvron2023llama}, Mixtral-8x22B \cite{jiang2024mixtral},  Mistral7B-v3 \cite{jiang2023mistral},  GPT4-Omni \cite{achiam2023gpt} and Gemini2-Flash \cite{team2023gemini}.
All experimental details can be found in Appendix \ref{app<code_hallucianation>: Experimental Details}.

% 4th category: comments and print statements 

% \begin{itemize}
%     \item Quantitative analysis: Please refer to the table from logs on date 11th June 2024
%     \item NER Insights -> Phantom vs ultra Phantom
%     \item Action Insights -> 
%     \item Code length size
%     \item Model Specific details 
%     \item Error Analysis 
%     \item Schema for results
% \end{itemize}

\begin{table*}[ht!]
    \centering
    \resizebox{\textwidth}{!}{

    \begin{tabular}{c|ccccc}
    \hline
    Models & Summary Length & CE count & Mapped $(\uparrow)$ & Unmapped $(\downarrow)$ & NL entities \\
    \hline

    Codellama-7B   & 236.10 & 8.638 & 80.17\% & 17.65\% & 2.18\% \\
    Mistral-7B     & 227.91 & 6.961 & 79.92\% & 17.22\% & 2.86\% \\
    Llama3-8B      & 257.21 & 9.45  & 84.50\% & 12.77\% & 2.73\% \\
    \hline
    Granite-20B    & 148.95 & 7.10  & 79.54\% & 19.64\% & 0.82\% \\
    Granite-34B    & 214.50 & 6.55 & 85.69\% & 12.41\% & 1.90\% \\
    Codellama-34B  & 278.67 & 8.22  & 79.78\% & 19.05\% & 1.17\% \\
    \hline
    Llama3-70B     & 313.58 & 10.01 & \textbf{88.44\%} & \textbf{8.69\%} & 2.87\% \\
    \end{tabular}
    }						
    \caption{Quantitative Analysis: This table presents summary statistics for all seven models. "CE count" is the number of code entities in the summary. "Mapped" is the percentage of code entities correctly mapped to the code, "Unmapped" refers to those unmapped, and "NL entities" refers to the remaining natural language (non-code) entities, all normalised by the total entity count.}
    \label{tab<code_hallucination>:Quantitative_Analysis}
\end{table*}

\textbf{Entity-Intent Verification:} In this aspect of evaluation, we aim to verify the intent of an individual entity. We report the results of entity-intent verification in the Table \ref{tab<code_hallucination>:hallucination_detection}. 
In general, we observe consistent improvements across all the models compared to the direct approach. 
It can be observed that the GPT4-Omni F1-Score is 0.73 while the Gemini F1 Score is 0.64. We observed close to random performance for other models like Llama3.1-70B, Llama3-8B and Mistral-7 B.
Upon analysis, we found that these models often classify INCORRECT tuples as CORRECT when the code references a function or library that is not defined in the input. Here, the model infers the functionality based on the library name (Identifier Name Bias \ref{sec:HC1-Name_Bias}), which can be difficult to verify. 

% \ref{tab<code_hallucination>:hallucination_detection}. It can be observed that the GPT4-Omni F1-Score is 0.61 while the Gemini F1 Score is 0.55. Upon analysis, we found that these models often classify INCORRECT tuples as CORRECT. This is mainly due to the convincing nature of the summary, which may also be subjective due to a lack of proper code context. For instance, we observe significant errors when the code references a function or library that is not defined in the input. In such cases, the model infers the functionality based on the library name (Identifier Name Bias \ref{sec:HC1-Name_Bias}), which can be difficult to verify.
% and upon analyzing the results, we identified that there is a scope for improvement in the relevant sentence identification method. In the case of generic entities such as datatypes and generic variable names such as \textit{property}, \textit{value}, etc., irrelevant sentences were identified as relevant, which contributed to the noise in entity-intent verification. 

\textbf{Instance Level Hallucination Verification:}
In this aspect of evaluation, we aim to verify the overall summary instance. To compare our approach, we consider a direct setup which involves providing a <code, summary> tuple to identify if the summary is hallucinated or not (Refer Appendix \ref{Appendix<Prompts>}, Figure \ref{box: Direct Intent Verification prompt}). We provide these results in Table \ref{tab<code_hallucination>:hallucination_detection}, and it can be observed that our approach provides significant improvement in F1-Score when compared to the Direct approach. In general, the direct evaluation method suffers from hallucinations, such as when identified entities for hallucination are absent from the summary or when natural language entities are mistakenly considered code entities, overall resulting in poor performance. This conveys that our finer-grained evaluation approach, which traces the entities from summary to code, provides a more reliable method to identify hallucinated summaries compared to traditional LLM-based approaches. 
% It also helps with interpretability as it identifies the hallucinated sections of the summary.  
% \begin{table}[h]
% \centering
% \begin{tabular}{l|ccc}
% \toprule
% & \textbf{P} & \textbf{R} & \textbf{F1} \\
% \midrule
% \textbf{GPT Direct} & 0.50 & 0.50 & 0.50\\
% \textbf{GPT ETF} & 0.79 & 0.89 & 0.82\\
% \end{tabular}
% \caption{GPT Instance Level Complete}
% \label{tab:classification_report_human_gpt_direct}
% \end{table}
% For this, we are considering the summaries from llama3-8B since the model is moderate in terms of (\%) unmapped entities highlighting that the model is neither too high nor too low in hallucination. 
\label{sec:analysis}
\section{Analysis}
In this section, we discuss various quantitative and qualitative insights of our framework, including its general predictive behaviour and cases of errors.

% 
% We first discuss summaries generated by individual models and then elaborate on the general predictive behaviour of our framework and cases of errors.  

% \textbf{Entity Extraction:}
% Some of the prominent confused class were Library and Method; Data Type and Function; Function and Algorithm; Library and File Name; Library and Data TYPE; CLASS and Function
% During entity extraction from the summary, we observed hallucinations in the predicted label set, where models tend to introduce extra labels not mentioned in the prompt but similar to the correct ones. For instance, the entity type "method" is often confused with "function." Among all code entity types, "VALUE" is the most frequently misclassified. This may be due to contextual ambiguity, as entities like "null" or numerical values—particularly their textual equivalents (e.g., "3" -> "three")—are often mistaken for other language entities.
% \textbf{Hallucination detection:}
\subsection{Quantitative Analysis}
% We report the quantitative analysis of obtained summaries in the Table \ref{tab<code_hallucination>:Quantitative_Analysis} and summarize our observations in this section. 
As shown in Table \ref{tab<code_hallucination>:Quantitative_Analysis}, Granite-20B produced shorter summaries, while Llama3-70B generated longer ones. Other models had similar average lengths, reflecting varying elaboration due to differences in training methodologies.
% \textbf{Fabricated Entities:} While performing named entity recognition, we observed that language models fabricated entities. This means there were entities recognized by the language model that were not present in the summary itself. These entities were directly filtered by string-matching heuristics and were discarded after the named entity step. 
% We observed that Llama3-70B had the least fabricated entities while Codellama-7B had the most fabricated entities, showing signs of high hallucination.
For entity mapping, we observe that Llama3-70B has the most mapped entities, indicating the tendency of the model to stay grounded. Granite-20B has the most unmapped entities, which indicates its tendency to produce content which may not be directly related to the code, leading to extrinsic hallucination.  

% For evaluating named entity recognition, we consider Jaccard Similarity in measuring the coverage of entities compared to ground truth. 
% We use the macro F1 score to evaluate the entity type classification of the NER. The results for NER can be found in the Table in \ref{tab<code_hallucination>:ner_results}.

\subsection{Taxonomy Analysis}
In this section, we discuss insights related to each category defined in the proposed taxonomy. As described in the Figure \ref{fig:hallucination_distribution}, we observe Code Complexity to be the most common hallucination cause. This arises due to the complicated variable names, which are not suggestive of what the code is about, the lengthy code or the complex logic. This may also stem from a lack of reasoning capability in models, often due to their small size or insufficient training. 
% (either the model can have limited reasoning capacity, or the code can be complex). 
Some of the most confused cases in our dataset involve function overloading and recursive functions, where the models often got confused between the caller function and the callee function. Other scenarios include:
\begin{inparaenum}[(1)]
\item the presence of multiple conditions within a boolean expression (e.g., an \texttt{if} condition with a long boolean argument containing multiple operators and operands),
\item multiple function calls within a single expression, and
\item the use of variable names that lack meaningful semantic interpretation.
\end{inparaenum}
% Other scenarios include 1) the presence of multiple conditions within a boolean expression (for example, an if condition with a long boolean argument that has multiple operators and operands), 2) multiple function calls within an expression and 3) cases where variable names used did not have any meaningful semantic interpretation. 

The next major category is Insufficient Knowledge, where models fail to correctly interpret fundamental aspects of the code or the libraries used. This includes cases where models cannot recognise function arguments or fail to understand library functionalities, often due to limited exposure during training or user-specific customisations of the libraries. Such hallucinations are more prevalent in smaller models, though occasional instances are also observed in larger models (greater than 30B).     

In addition to the above categories, we also observe notable instances of Name Bias and hallucination influenced by Natural Language Context. Primarily, Name Bias occurs when the models rely too much on the semantic interpretation of the identifier names. This often overlaps with Insufficient Knowledge, where the model lacks grounding but guesses based on naming, and can be attributed to inadequate model training. For example, there might be a line which mentions a function name not defined in the given code itself; in such scenarios, models are `forced' to rely on the semantics. This is also the reason that there are some frequent cases of Insufficient knowledge and Name Bias together. It is noteworthy to mention that almost all the annotators also expressed ambiguity in such instances, making it difficult to annotate even for humans. Hallucination due to natural language context, like comments or log statements, occurs when the comments represent outdated information or when comments discuss some additional functionality not suggested in the code logic.          

\subsection{Predictive Analysis}

\begin{tcolorbox}[,
    colback=gray!10,     % Background color
    colframe=gray!30,    % Border color
    coltitle=black,
    fonttitle=\bfseries, % Font style for the title
    top=0cm,
    left=0cm,            % Left margin
    right=0cm,           % Right margin
    bottom=-2mm-1.5mm,
    float=ht!
    ]
\small
\begin{lstlisting}
private List<Transaction> retrieveTransactions(String rowStatusCd) throws HubException{............}
\end{lstlisting}
\textbf{Summary}:
This method retrieves transactions using the \textcolor{red}{``Hub" API} based on the input parameters by the \textcolor{red}{``Hub database"}. It interacts with several key entities, including \textcolor{red}{``squid:s1166"} and \textcolor{red}{``squid:s1172"}, to gather the data......
\tcbsubtitle[halign=center]{Analysis: Hallucination and Unmapped Entities}
\label{box: Analysis: Hallucination due to Unmapped Entities}
\end{tcolorbox}

Our framework captures ungrounded entities in summaries. For example, when the model refers to a non-existent Hub API based on the keyword `HubException", the unmapped entities such as "Hub" are identified. This enables fine-grained and interpretable detection of hallucinations. 
% We can see that the ungrounded entities like "Hub" are captured here, leading to more fine-grained and interpretable detection of hallucination.

\subsection{Error Analysis}
This section discusses the two major error cases which can be observed in our framework: 

\noindent
\textbf{Error Case 1-Creative Summary:} 
Models sometimes add elaborative details that may be accurate but not explicitly present in the code. In the given example, the code summary discusses a more elaborate version of the input code by restating an elongated version. Here, the entities present in the summary are predicted to be ungrounded by our framework.

\begin{tcolorbox}[
    colback=gray!10,     % Background color
    colframe=gray!30,    % Border color
    coltitle=black,
    fonttitle=\bfseries, % Font style for the title
    top=0mm,
    left=0cm,            % Left margin
    right=0cm,           % Right margin
    bottom=-2mm-1.5mm,
]
\small
\begin{lstlisting}
public static int writeShort(ArrayView target, int offset, short value) 
    {
    return writeShort(target.array(), target.arrayOffset()+offset, value);
  }
\end{lstlisting}

\textbf{Summary}:
\
Here's a more detailed look at the implementation:
....\{JAVA CODE\}.....

\tcbsubtitle[halign=center]{Creative Summary by Mistral 7B}
\label{box: Error-Analysis: Creative Summary Generation}
\end{tcolorbox}

\textbf{Error Case 2- Changed Entity Form:} Language models may not use the exact names of code entities in summaries. For example, the entity "PreparedStatement" may be referred to as "prepared statement," which could be missed during named entity recognition due to the change in form. The verification of these kinds of summaries may not reflect the inaccuracies due to such sentences.

\begin{tcolorbox}[
    colback=gray!10,     % Background color
    colframe=gray!30,    % Border color
    coltitle=black,
    fonttitle=\bfseries, % Font style for the title
    top=0mm,
    left=0.1cm,            % Left margin
    right=0.1cm,           % Right margin
    bottom=-2mm-1.5mm,
]
\small
\begin{lstlisting}
protected PreparedStatement setObjects(Connection connection, String query, Object... objects) throws SQLException {
    PreparedStatement ps = statement(connection, query);
    setObjects(ps, objects);
    return ps;
        }
\end{lstlisting}

% \begin{lstlisting}
% protected PreparedStatement setObjects(Connection connection, String query, Object... objects) throws SQLException {....}
% \end{lstlisting}

\textbf{Summary}:
\
......The method is used to create and execute a \textbf{prepared statement} using the given connection and query. The method takes an array of objects as input, which are used to set the parameters in the \textbf{prepared statement}. The method returns the \textbf{prepared statement} object that was created and executed......

\tcbsubtitle[halign=center]{ Changed Entity Form by Codellama 7B}
\label{box: Error-Analysis: Changed Entity by LLama3-70B}
\end{tcolorbox}

\section{Conclusion and Future Work}
% In this work, we discussed the challenges of detecting hallucinations in generated code summaries. We then proposed an entity-focused intent verification technique for detecting hallucinations in code summaries. In future, this framework can be enhanced by incorporating a multi-agent system, leveraging multiple LLMs in tandem to improve prediction accuracy. Additionally, the current framework can be further developed to better mitigate the occurrence of hallucinations.

In this work, we address the problem of detecting hallucinations in code summarisation, a task that demands a deep understanding of both programming and natural languages. By introducing a novel dataset, CodeSumEval and an Entity Tracing Framework (ETF), we present a promising approach to grounding code entities within summaries. This enables a more explainable and accurate evaluation of code summaries. In the future, we plan to explore agent collaboration where different LLMs verify entities and intents independently. Further improvements may involve fuzzy or embedding-based entity detection, documentation lookup or extension to mitigate hallucinations.

\textbf{\section{Limitations}}
While the framework is designed to be generic, certain components, such as code parsers for entity detection, may be unavailable for low-resource programming languages like COBOL or Perl. This limits its applicability to low-resource programming languages (e.g., COBOL, Perl) lacking robust static analysis tools. Further, the entity detection and verification stages depend heavily on large-scale LLMs. Given the complexity of this task, smaller open-source models may struggle to perform effectively, reinforcing the need for larger LLMs. This may not be scalable and often demands significantly greater computational resources. Finally, our multi-stage verification process involves multiple LLM calls, making real-time deployment and integration into production-level developer tools challenging without further optimisation.

\bibliography{acl_latex}

\appendix

\label{sec:appendix}
% We organize the appendix to cover the following sections:
% \begin{itemize}
%     \item Prompts - Discusses the different prompts we used for summary generation, NER, intent verification tasks.
%     \item Experimental Setup - Captures the details of the hardware setup and model hyper parameters
% \end{itemize}

\section{Prompts}
\label{Appendix<Prompts>}

\label{box: summary generation prompt}
\begin{tcolorbox}[title=Summary Generation Prompt,
    colback=gray!10,     % Background color
    colframe=gray!30,    % Border color
    coltitle=black,
    fonttitle=\bfseries, % Font style for the title
    left=0.2cm,            % Left margin
    right=0.2cm,           % Right margin
]

Assume you are an expert in understanding JAVA code. \\
Question: As a Java Expert, please provide a detailed summary of the following Java code with the following sections: \\
1. Inputs and outputs of the method \\
2. Business purpose \\
3. Detailed functional summary of the method. \\
``` \\
 \{CODE\} \\
```
\end{tcolorbox}
\captionof{figure}{Summary Generation Prompt- This prompt was used for generating the summaries from different language models}

\begin{tcolorbox}[title=Intent Verification Prompt,
    colback=gray!10,     % Background color
    colframe=gray!30,    % Border color
    coltitle=black,
    fonttitle=\bfseries, % Font style for the title
    left=0.1cm,            % Left margin
    right=0.1cm,           % Right margin
]
Assume you are an expert in understanding JAVA code. Your task is to verify whether the description of '{mapped\_entity}' in the given text is correct, incorrect, or irrelevant with respect to the code.
Only output one of the following labels: [``CORRECT", ``INCORRECT", ``IRRELEVANT"].\\
Description:
\\
$\{relevant\_sent\}$\\
$[CODE]$\\
\{CODE\}\\
$[\mathbin{/} CODE]$
\end{tcolorbox}
\captionof{figure}{Intent Verification Prompt- This prompt was used for verifying the description of a given entity based on the sentences that mention the entity}\label{box: Intent Verification prompt}

\label{box: named entity recognition prompt}

\begin{tcolorbox}[title=Named Entity Recognition Prompt,
    colback=gray!10,     % Background color
    colframe=gray!30,    % Border color
    coltitle=black,
    fonttitle=\bfseries, % Font style for the title
    left=0.1cm,            % Left margin
    right=0.1cm,           % Right margin
]
Assume you are an expert in understanding Java and performing named entity recognition related to Java code. You have to label the entities by considering the following labels:\\
\\
Code Entities: CLASS, VARIABLE, FUNCTION, LIBRARY, VALUE, DATA TYPE, and HTML or XML TAG\\
Natural Language Entities: APPLICATION, UI ELEMENT, LANGUAGE, DATA STRUCTURE, ALGORITHM, FILE TYPE, FILE NAME, VERSION, DEVICE, OS, WEBSITE, and USER NAME.\\

For every entity in the input, mention the entity\_type in the given format only. Strictly follow this template and only print the output without any other words. You can follow the example below:
\\
``` \\
 \{Incontext Example\} \\
```

Now consider the summary describing the code below:\\
 \{generated\_summary\} 
\end{tcolorbox}
\captionof{figure}{Named Entity Recognition Prompt}

\begin{tcolorbox}[title=Direct Evaluation Prompt,
    colback=gray!10,     % Background color
    colframe=gray!30,    % Border color
    coltitle=black,
    fonttitle=\bfseries, % Font style for the title
    left=0.1cm,            % Left margin
    right=0.1cm,           % Right margin
]
Assume you are an expert in understanding JAVA code. Your task is to verify if the description of the code entities present in the given summary is correctly described or NOT as per the code logic.  
Output all the `entity\_name' and a relevant\_sentence' corresponding to the `entity\_name', which are incorrectly described. Do not provide any other details.
Strictly follow this format: $[{entity\_name:``", relevant\_sentence:``"}]$

Summary:\\
 \{SUMMARY\}\\
 
Code:\\
 \{CODE\} 
\end{tcolorbox}
\captionof{figure}{Direct Evaluation Prompt- This prompt was used to detect the hallucinated entities and sentences from the summary without breaking into entities}\label{box: Direct Intent Verification prompt}

% \section{Dataset}

\section{Experimental Setup}
\label{app<code_hallucianation>: Experimental Details}

%% We need to say that we have studied on different sizes and family of the model. We have to talk about how are they different in terms of architecture and design. 
%Mentioning why we choose instruct model can also be a good idea.

% \begin{itemize}
%     \item LLama 3-8B-Instruct 
%   \item LLama 3-70B-Instruct  
%   \item Granite-20b-Code-Instruct  
%   \item Mistral 7B v2-Instruct
%   \item CodeLlama-7b-Instruct  
%   \item CodeLlama-34b-Instruct 
%   \item Granite 34b-Code-Instruct  
% \end{itemize}
In our setup, we conducted all the experiments using NVIDIA A100-SXM4-80GB GPU in a single or multi-GPU environment. For our experiments, we consider instruction-tuned versions of the SOTA code and language models,  from IBM-Granite family (20B-instruct; 34B-instruct) \cite{mishra2024granite}, Llama3 family (8B-instruct and 70B-instruct) \cite{touvron2023llama}, CodeLlama family (7B and 34B) \cite{roziere2023code} and Mistral family (7B-instruct) \cite{jiang2023mistral}. We use the GPT4-Omni version for our framework and keep the temperature at 0.3 and set max\_new\_tokens to 4000.

\section{NER Evaluation}
\label{Appendix<ner_evaluation>}
This section discusses the NER performance of various models considered in this work. 
To perform NER using LLMs, we provide the code summary and NER tagset in the prompt (Appendix \ref{box: named entity recognition prompt}) using a one-shot in-context example to extract all the entities discussed in the summary, accompanied by their types. 
To evaluate the entity extraction, we assess two key aspects: entity coverage and entity type correctness. 1) \textbf{Entity Coverage}: This measures whether all valid entities in the summary are detected. We quantify this using the Jaccard Similarity between the entities in the generated output and those in the ground truth. 2) \textbf{Entity Type Correctness}: This evaluates whether the detected entities have been assigned the correct types. For this, we use the F1 score as the metric.

\begin{table}[ht!]
    \centering
    % \small
    \begin{tabular}{c|cccc}
    \hline
     Models &  Jaccard Similarity & F1 
\\
    \hline
 
    GPT-4-Omni &  0.81  &   0.92   \\
    Gemini-1.5-Flash&  0.64 & 0.92      \\
    \hline

    \end{tabular}
								
    \caption{\textbf{NER Results on Human Data}}
    % Here, the term JS indicates ``Jaccard Similarity", which indicates the coverage capabilities of the language model}
    \label{tab<code_hallucination>:ner_results_human}

\end{table}

We observed a good correlation between GPT4-Omni and human data and, therefore, used it for NER in our pipeline. As an additional contribution, we also evaluate the open-source models considered in this work for the task of Named Entity Recognition on summaries generated from 600 code snippets initially sampled from CodeXGlue data using GPT predictions as ground truth.

\begin{table}[ht!]
    \centering
\resizebox{0.48\textwidth}{!}{
    \begin{tabular}{c|ccc}
    \hline
    Models & JS & F1 & Fabricated $(\downarrow)$ \\
    \hline
    Codellama-7B     & 0.4586  & 0.84 & \textit{35.68\%} \\
    Mistral-7B       & 0.4458  & 0.65 & 8.59\% \\
    Llama3-8B        & 0.5298  & 0.78 & 6.44\% \\
    \hline
    Granite-20B      & 0.4897  & 0.85 & 4.25\% \\
    Granite-34B      & 0.48181 & 0.84 & 2.45\% \\
    Codellama-34B    & 0.5079  & 0.83 & 4.87\% \\
    Llama3-70B       & \textbf{0.5981} & \textbf{0.90} & \textbf{0.15\%} \\
    \hline
    \end{tabular}
    }				
    \caption{\textbf{NER Results on GPT Data}. Here, `JS' refers to Jaccard Similarity, and `Fabricated' refers to the percentage of fabricated entities identified during named entity recognition, normalised by the total entities as mentioned in Table \ref{tab<code_hallucination>:Quantitative_Analysis}.}
    % Here, the term JS indicates ``Jaccard Similarity", which indicates the coverage capabilities of the language model}
    \label{tab<code_hallucination>:ner_results}

\end{table}

% \textbf{Fabricated Entities:} While performing named entity recognition, we observed that language models fabricated entities. This means there were entities recognized by the language model that were not present in the summary itself. These entities were directly filtered by string-matching heuristics and were discarded after the named entity step. 
% We observed that Llama3-70B had the least fabricated entities while Codellama-7B had the most fabricated entities, showing signs of high hallucination.

\section{Intent Detection}
\label{app:intent_detection_details}
In this section, we describe the two distinct approaches for intent detection. 
\subsection{String Matching Heuristics}
\label{app:string-matching-heuristics}
By string matching heuristics, we mean character-level matching with the following regex expressions:
\begin{itemize}
    \item The word is either preceded by or succeeded by a space char
    \item ignore the “\verb|`|” characters since some of the entities are enclosed using these quoted marks by models.
    \item Account for brackets: some of the function names in the summary include “()” and some of the variables include “[]”
\end{itemize}

The above regexes are designed to capture all the cases of entity forms in summary, evaluated in a single module.

\subsection{Prompt based Approaches}
Here, we discuss the general prompt-based approach we tried for Intent detection. We give the complete prompt in Figure \ref{box: Intent Detection prompt}. Qualitatively, we observed the following drawbacks of this approach:

\begin{itemize}
    \item We observe high Hallucination in the generated output, which leads to the introduction of fabricated sentences not present in the summaries. 
    \item We observed inaccurate extraction of the sentences, where the extracted sentence has slight variations from the original sentence present in the summary.

    \item We observe missing sentences, i.e., not all the sentences discussed in the summary are captured in the generated output.

\end{itemize}

\begin{tcolorbox}[title=Intent Detection Prompt,
    colback=gray!10,     % Background color
    colframe=gray!30,    % Border color
    coltitle=black,
    fonttitle=\bfseries, % Font style for the title
    left=0.1cm,            % Left margin
    right=1cm,           % Right margin
]
Assume you are a Java expert. You have to identify all the relevant sentences about the given entity. Here, a sentence is relevant to the mapped entity if the sentence discusses the given entity. You have to generate the output strictly in the JSON format: $\{{entity\_name:``", relevant\_sentence:``"}\}$\\

Given Entity: \\
 \{mapped entity\}\\

Given Summary:\\
 \{SUMMARY\}
\end{tcolorbox}
\captionof{figure}{Intent Detection Prompt- This prompt was used to retrieve all the relevant sentences from the summary}\label{box: Intent Detection prompt}

These observations led us to prefer simpler string-matching heuristics, which are significantly cheaper in computational aspects.

\section{Annotation Details}
We discuss our annotation process and the annotator's guidelines here: 
\subsection{Background}
The dataset was annotated by eight annotators who are experts in Java and held at least a Master’s degree in Computer Science, with some having a PhD in the field. On average, the annotators had 4+ years of experience in Java programming.

\subsection{Guidelines}
The annotation process was conducted in two stages. In the first stage, we implemented a three-step procedure to annotate hallucinations in code summaries independent of specific hallucination categories. The first step of the annotation process involved validating the Named Entity Recognition Output. This involved annotating missed or incorrectly identified entities in the summary itself by selecting the appropriate label from the drop-down.

The second step of the annotation involved evaluating if each sentence accurately describes the entity in the code snippet, marking the entity-sentence pair as:
\begin{itemize}
    \item \textbf{CORRECT:} If the relevant sentence correctly describes the code
    \item \textbf{INCORRECT:} If the relevant sentence incorrectly describes the code
    \item \textbf{IRRELEVANT:} If the relevant sentence does not talk about the mapped entity itself
\end{itemize}

The third step involved rating the summary based on hallucination severity as 
\begin{itemize}
    \item \textbf{POOR} (Most part is hallucinated): The generated code summary shows below-average correctness.
    \item \textbf{FAIR} (Only some part is hallucinated): The generated code summary meets expectations.
    \item \textbf{GOOD} (Almost no hallucination): The generated code summary is completely correct.
\end{itemize}
\label{appendix:annotation_guidelines} 

The second stage involved defining hallucination categories based on annotator feedback and organising them into a structured taxonomy (Figure \ref{fig:taxonomy-code-hallucination}). This finalised taxonomy was then provided to the annotators, who were asked to assign a specific hallucination category from the predefined options. Annotators were also encouraged to include comments explaining their annotations, as these explanations can be useful for researchers utilising our dataset.

\end{document}